\documentstyle[12pt]{article}
\def\ie{{\em i.e.}}
\def\eg{{\em e.g.}}

\def\beq{\begin{equation}}
\def\eeq{\end{equation}}

\catcode`\@=11 % This allows us to modify PLAIN macros.

\def\lsim{\mathrel{\mathpalette\@versim<}}
\def\gsim{\mathrel{\mathpalette\@versim>}}
\def\@versim#1#2{\vcenter{\offinterlineskip
    \ialign{$\m@th#1\hfil##\hfil$\crcr#2\crcr\sim\crcr } }}
\def\etal{{\em et. al.}}
\def\JL{J. L. Lopez}
\def\DVN{D. V. Nanopoulos}
\def\AZ{A. Zichichi}

\def\r#1{$\bf#1$}
\def\rb#1{$\bf\overline{#1}$}

\def\t1{{\tilde 1}}

\def\GeV{\,{\rm GeV}}

\def\to{\rightarrow}
\def\pb{\,{\rm pb}}
\def\ipb{\,{\rm pb}^{-1}}

\def\NPB#1#2#3{Nucl. Phys. B {\bf#1} (19#2) #3}
\def\PLB#1#2#3{Phys. Lett. B {\bf#1} (19#2) #3}

\def\PRD#1#2#3{Phys. Rev. D {\bf#1} (19#2) #3}

\def\MODA#1#2#3{Mod. Phys. Lett. A {\bf#1} (19#2) #3}
\def\IJMP#1#2#3{Int. J. Mod. Phys. A {\bf#1} (19#2) #3}
\def\TAMU#1{Texas A \& M University preprint CTP-TAMU-#1}

\begin{document}
% TH format
\begin{flushright}
\baselineskip=12pt
{CTP-TAMU-06/95}\\
{ACT-02/95}\\
{hep-ph/9502414}\\
\end{flushright}

\begin{center}
%\vglue 0.5cm
{\Huge\bf A string no-scale supergravity model and its experimental
consequences\\}
\vglue 1cm
{JORGE L. LOPEZ$^{1,2}$, D. V. NANOPOULOS$^{1,2}$, and A.~ZICHICHI$^{3}$\\}
\vglue 0.4cm
{\em $^{1}$Center for Theoretical Physics, Department of Physics, Texas A\&M
University\\}
{\em College Station, TX 77843--4242, USA\\}
{\em $^{2}$Astroparticle Physics Group, Houston Advanced Research Center
(HARC)\\}
{\em The Mitchell Campus, The Woodlands, TX 77381, USA\\}
{\em $^{3}$CERN, 1211 Geneva 23, Switzerland\\}
\baselineskip=12pt
\end{center}

\vglue 0.5cm
\begin{abstract}
We propose a string-derived model based on the gauge group $SU(5)\times U(1)$
which satisfies the stringent constraints from no-scale supergravity, allows
gauge coupling unification at the string scale, and entails previously
unexplored correlations among various sectors of the model. All supersymmetric
observables are given in terms of a single mass parameter with self-consistency
of the model determining the rest, including $\tan\beta=2.2-2.3$ and
$m_t\approx175\GeV$. A small non-universality of the scalar masses at the
string scale produces a downward shift in the right-handed slepton masses at
the electroweak scale, such that for $m_{1/2}\gsim180\GeV$ these particles
become lighter than the lightest neutralino. This cutoff in the parameter space
entails the imminent discovery of charginos at the Tevatron via trilepton
events ($m_{\chi^\pm_1}<90\GeV$). Also, the lightest Higgs boson
($m_h<90\GeV$), the lightest chargino, and the right-handed sleptons
($m_{\tilde\ell_R}<50\GeV$) should be readily observable at LEPII. We also
discuss the model predictions for $B(b\to s\gamma)$, $(g-2)_\mu$, $R_b$, and
the prospects for direct neutralino dark matter detection.
\end{abstract}

\vspace{0.5cm}
% TH format
\begin{flushleft}
\baselineskip=12pt
{CTP-TAMU-06/95}\\
{ACT-02/95}\\
February 1995
\end{flushleft}
\newpage

\setcounter{page}{1}
\pagestyle{plain}
\baselineskip=14pt

Despite all the experimental evidence in support of the Standard Model
of the strong and electroweak interactions, many physicists believe that
it must be extended so that its many ad-hoc parameters may find explanation
in a more fundamental theory. Among the various avenues that lead away from
the Standard Model, the ideas of supersymmetry, supergravity, and superstrings
are particularly compelling in tackling the shortcomings of the Standard
Model. Low-energy supersymmetry predicts the existence of a superpartner for
each of the Standard Model particles with well determined interactions but
undetermined supersymmetry-breaking masses, although these should not exceed
the TeV scale if the gauge hierarchy problem is to remain at a tolerable level.
Supergravity provides an effective theory of supersymmetry breaking in terms of
two input functions, the K\"ahler function and the gauge kinetic function. With
these inputs all supersymmetry-breaking masses can be calculated in terms of a
single parameter: the gravitino mass ($m_{3/2}$). Superstrings provide the
final link by allowing a first-principles calculation of these two input
functions in any given string model, therefore having a single parameter
effectively describing the physics of supersymmetry breaking.  At low-energies
a new parameter arises, namely the ratio of vacuum expectation values
($\tan\beta$) of the two Higgs-boson doublets minimally required in
supersymmetric models. However, minimization of the electroweak
scalar potential with respect to the two neutral Higgs fields provides two
additional constraints which effectively reduce the number of parameters to
zero, and a {\em no-parameter} model is obtained.

In this Letter we describe one such no-parameter model obtained in the context
of string no-scale supergravity \cite{LN,Ferrara}. In contrast with traditional
unified models with ad-hoc ``string-inspired" choices for the supersymmetry
breaking parameters, in our string-derived model the parameters describing
the various sectors of the model (\ie, gauge group, matter spectrum,
superpotential, and supersymmetry breaking) are calculated from first
principles. Our model is also consistent with the postulates of no-scale
supergravity that open the way for a dynamical determination of all mass scales
(\eg, $m_{3/2}$ and $M_Z$), which must otherwise be self-consistently or
experimentally determined. The existence of such model is particularly
remarkable given the strong restrictions that no-scale supergravity imposes on
string model-building. In practice we are unable to extract all of the in
principle available string information, and thus our model in fact has one free
parameter (\ie, $m_{3/2}$ or the mass scale of the supersymmetric spectrum).
However, self-consistency constraints of the model  strongly restrict the
allowed range of this one parameter, and the same holds for the various
experimental predictions of the model.

Since there are so many possible string models, we guide our search for a
realistic model by a few principles:
(i) a unified gauge group which can break down to the Standard Model gauge
group, (ii) a matter content which reduces to the supersymmetric Standard Model
at low energies and that allows unification of the gauge couplings at the
string scale ($M_U\sim10^{18}\GeV$), and (iii) a low-energy effective theory
with the no-scale supergravity structure with vanishing vacuum energy
\cite{Cremmer,Lahanas,LNrev}.

A string model satisfying the first two constraints was derived in
Ref.~\cite{search}. This model has the observable sector gauge group
$SU(5)\times U(1)$ \cite{flipped}, three generations of quark and lepton
superfields, and two light Higgs doublet superfields. The
unified gauge symmetry is broken down to the Standard Model gauge group via
vacuum expectation values of scalar Higgs fields in \r{10},\rb{10}
representations. Moreover, the gauge symmetry constraints entail superpotential
interactions which suppress naturally dangerous dimension-five proton decay
operators, provide an elegant solution to the doublet-triplet splitting
problem, and a novel see-saw mechanism for neutrino masses \cite{Moscow}.
The model also predicts the existence of intermediate-scale vector-like
particles ($Q,\bar Q$ and $D^c,\bar D^c$) contained in one set of
\r{10},\rb{10} representations, with masses consistent \cite{search} with those
that allow unification of the gauge couplings at the string scale
(\ie, $m_Q\sim10^{12}\GeV$, $m_{D^c}\sim10^6\GeV$ \cite{LNZI}) .

In Ref.~\cite{LN} a study was performed of the constraints imposed by no-scale
supergravity on free-fermionic string model-building. No-scale supergravity
requires particular forms of the K\"ahler function such that the vacuum energy
vanishes and the potential possesses flat directions which leave the gravitino
mass undetermined \cite{Cremmer}. Minimization of the electroweak-scale scalar
potential with respect to the gravitino mass (the no-scale mechanism) then
determines its value \cite{Lahanas}. This mechanism becomes unstable to loop
corrections unless the quantity ${\rm Str}\,{\cal M}^2=2Qm^2_{3/2}$ (a weighted
sum of scalar and fermion supersymmetry-breaking masses) vanishes at the scale
of supersymmetry breaking. In Ref.~\cite{LN} it was shown that the string model
derived in Ref.~\cite{search} possesses a K\"ahler function that depends on a
{\em single} modulus field ($\tau$), besides the dilaton ($S$), with vanishing
vacuum energy
(the goldstino field is $\tilde\eta\propto S+\sqrt{2}\,\tau$) and with the
desired flat direction of the scalar potential. Moreover, the quantity $Q$ was
shown to be sufficiently small in first approximation, and plausibly vanishing
in a complete (although impracticable) calculation. We should remark
that the constraints from string no-scale supergravity are not satisfied
automatically. In fact, most of the string models explored in Ref.~\cite{LN}
did not satisfy them. Also, the dynamical determination of $m_{3/2}$ via the
no-scale mechanism is at the moment hampered by a new uncalculated parameter
quantifying a remnant vacuum energy at high energies~\cite{KPZ}.

In our string no-scale supergravity model it is possible to compute all the
soft-supersymmetry-breaking parameters at the string scale in terms of
$m_{3/2}$ \cite{LN}:
\begin{itemize}
\item Gaugino masses (universal): $m_{1/2}=m_{3/2}$
\item Scalar masses:
\begin{itemize}
\item First generation: $m^2_{Q_1,U_1^c,D_1^c,L_1,E_1^c}=0$
\item Second generation: $m^2_{Q_2,U_2^c,D_2^c,L_2,E_2^c}=0$
\item Third generation: $m^2_{Q_3,D^c_3}=m^2_{3/2}$, $m^2_{U^c_3,L_3,E^c_3}=0$
\item Higgs masses: $m^2_{H_1}=m^2_{H_2}=0$
\end{itemize}
\item Trilinear scalar couplings (universal): $A=m_{3/2}$
\item Bilinear scalar coupling: $B=m_{3/2}$
\end{itemize}
In addition, the parameters in the superpotential have been calculated
\cite{search}. Among these one finds the Higgs mixing term $\mu H_1 H_2$, which
arises as an effective coupling at the quintic level in superpotential
interactions, and gives rise to the $B$ parameter quoted above \cite{LN}.
Our present inability to reliably estimate the value of $\mu$ makes this the
single parameter of the model.

Another important superpotential coupling is the top-quark Yukawa coupling,
which in Ref.~\cite{search} was originally found to be $\lambda_t=g\sqrt{2}$,
where $g\approx0.83$ is the unified gauge coupling at the string scale,
obtained by running up to the string scale the Standard Model gauge couplings
\cite{LNZI}. The properly normalized top-quark Yukawa coupling is however found
to be $\widehat\lambda_t=g^2$, once a recently derived normalization factor is
inserted \cite{LN}. The top-quark mass itself cannot be yet determined since it
also depends on the low-energy parameter $\tan\beta$ (\ie,
$m_t\propto\sin\beta$).

The low-energy theory, obtained by renormalization group evolution from the
string scale down to the electroweak scale, thus depends on only one parameter
($m_{3/2}$ or $m_{1/2}$) since the magnitude of the Higgs mixing term $|\mu|$
and $\tan\beta$ can be self-consistently determined from the minimization of
the one-loop electroweak effective potential. Moreover, we find that the
constraint $B=m_{3/2}$ can only be satisfied for $\mu<0$. This general
procedure has been carried out before in supergravity models \cite{SugraCalcs},
however, with the further specification of $B$, the numerical computations
which determine the value of $\tan\beta$ become rather elaborate
\cite{LNZI,LNZII}. In the present
case a novelty arises because the scalar masses given above are not universal
at the string scale. This non-universality entails a modification of the usual
renormalization group equations \cite{LM}, which amounts to shifts in the
squared scalar mass parameters at low energies: $\Delta m^2_i=-c^2Y_if$, where
$Y_i$ is the hypercharge,
\begin{equation}
c^2=m^2_{H_2}-m^2_{H_1}+\sum_{i=1,2,3}\left(m^2_{Q_i}+m^2_{D^c_i}+m^2_{E^c_i}
-m^2_{L_i}-2m^2_{U^c_i}\right)=2m^2_{1/2}\ ,
\label{c^2}
\end{equation}
is the non-universality coefficient at the string scale, and $f\approx0.060$
is an RGE coefficient \cite{LM}. These shifts are most significant for the
right-handed sleptons ($\tilde\ell_R=\tilde e_R,\tilde\mu_R,\tilde\tau_R$)
whose masses are
\begin{equation}
m_{\tilde\ell_R}^2=a\,m^2_{1/2}+\tan^2\theta_W
M^2_W(\tan^2\beta-1)/(\tan^2\beta+1)\ ,
\label{mellR}
\end{equation}
with $a=0.153$ in the usual universal case, but $a=0.153-0.120=0.034$ in our
non-universal case. For the other scalars the (usual) coefficient $a$ is much
larger and the effect of the shift (${\cal O}(0.1)$) is relatively small. The
significance of the downward shift on $m_{\tilde\ell_R}$ relates to the
lightest supersymmetric particle, which is stable and should be neutral and
colorless \cite{EHNOS}. For the lowest allowed values of $m_{1/2}$, this
particle is the lightest neutralino $\chi^0_1$ with
$m_{\chi^0_1}\approx0.25m_{1/2}$. From Eq.~(\ref{mellR}) we see that as
$m_{1/2}$ increases there is a critical value
$m_{1/2}^{\rm c}$ above which the $\tilde\ell_R$ become the lightest
supersymmetric particles. Since this is phenomenologically unacceptable, we
cutoff the single parameter of the model at this critical value. This cutoff
turns out to be rather restrictive. Another novelty in our model is that the
top-quark mass is self-consistently determined by the value of $\tan\beta$
which results from the various other constraints.

\begin{figure}[p]
\vspace{6in}
\includegraphics{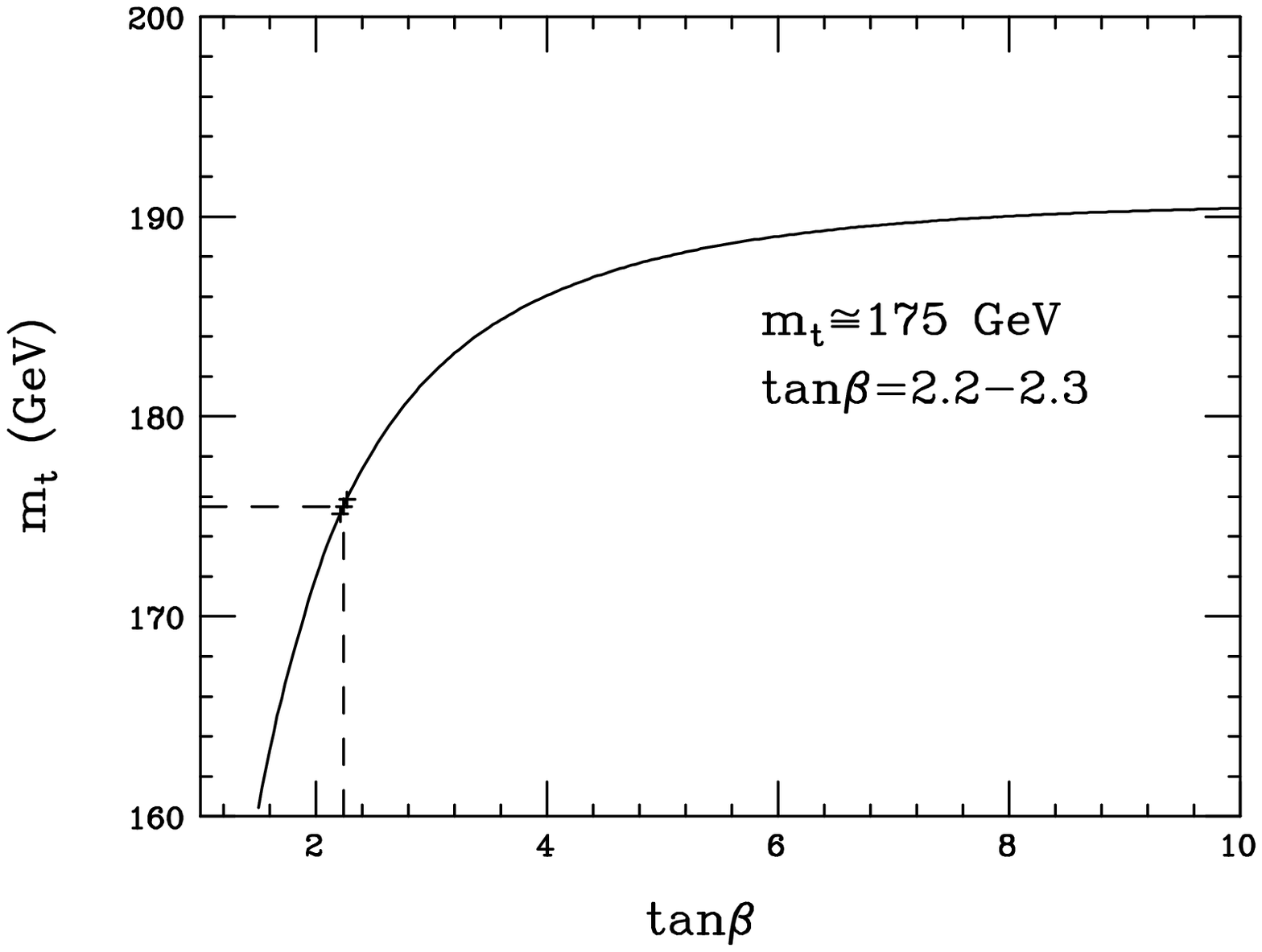}
\caption{The top-quark mass versus $\tan\beta$. Self-consistency of the
model requires $\tan\beta=2.2-2.3$ and thus $m_t\approx175\,{\rm GeV}$.}
\label{t-zero}
\end{figure}

\begin{figure}[p]
\vspace{6in}
\includegraphics{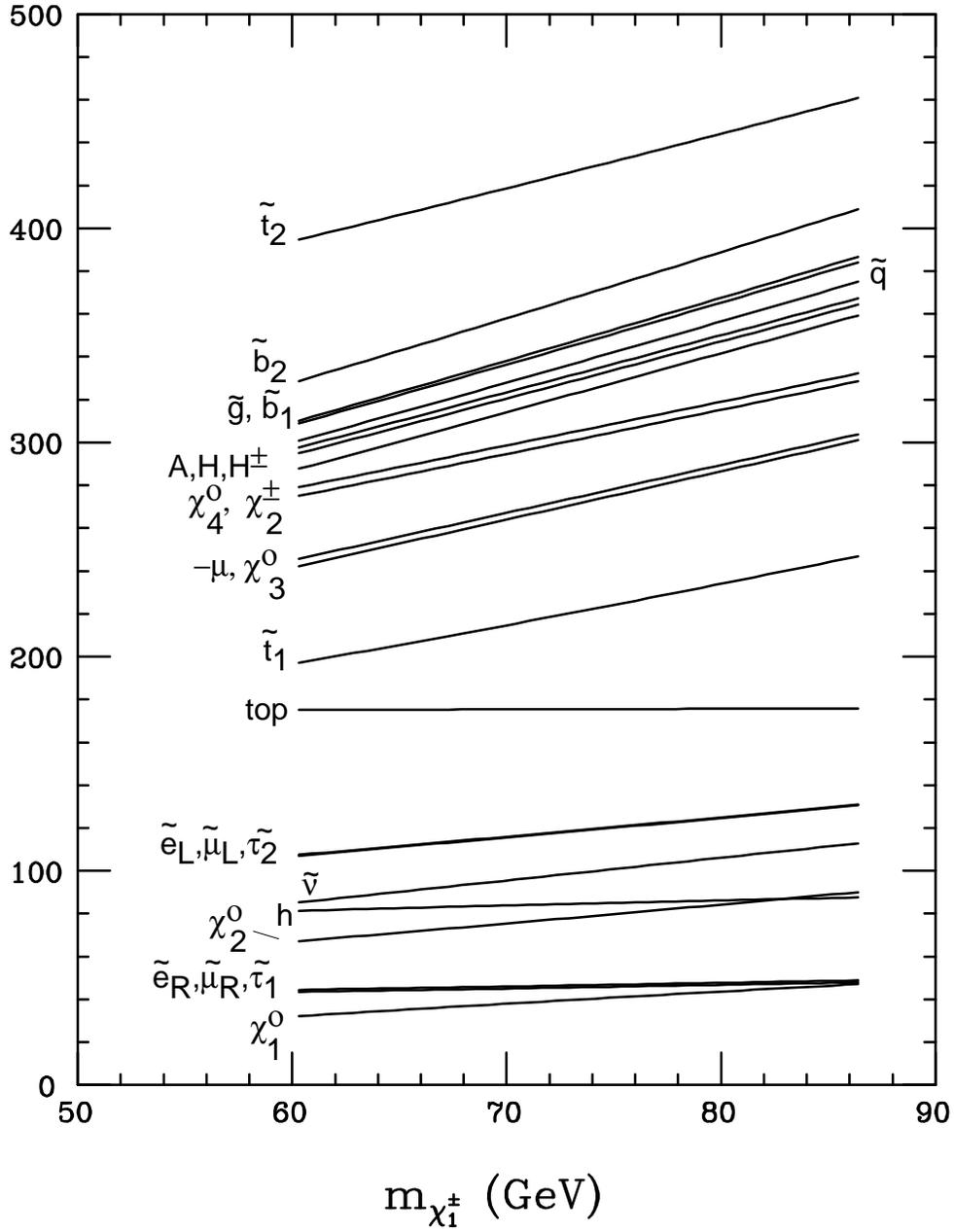}
\vspace{1cm}
\caption{The full sparticle and Higgs-boson mass spectrum versus the chargino
mass. Here $m_{\tilde q}$ is the average first- and second-generation squark
mass. Note that the spectrum cuts off when $m_{\tilde
e_R,\tilde\mu_R,\tilde\tau_1}=m_{\chi^0_1}$.}
\label{Spectrum}
\end{figure}

Once the calculations described above are performed we find $m_{1/2}^{\rm
c}\approx180\GeV$ and $\tan\beta=2.2-2.3$. The latter result allows a precise
determination of the (``pole") top-quark mass
\begin{equation}
m_t\approx175\GeV\ ,
\label{mt}
\end{equation}
which depends on $\sin\beta=\tan\beta/\sqrt{1+\tan^2\beta}$, as shown in
Fig.~\ref{t-zero}. With such small value of $\tan\beta$, the bottom and tau
Yukawa couplings at the string scale should be comparable
$\lambda_b\sim\lambda_\tau\sim0.01$ \cite{t-paper}. This matter will be
addressed in the context of this model elsewhere. The full mass spectrum is
shown in Fig.~\ref{Spectrum}. Note that the right-handed sleptons
($m_{\tilde\ell_R}<50\GeV$), the lightest chargino ($m_{\chi^\pm_1}<90\GeV$),
and the lightest Higgs boson ($m_h<90\GeV$), should all be within the reach of
the Tevatron or LEPII, as we now discuss.

\begin{figure}[p]
\vspace{6in}
\includegraphics{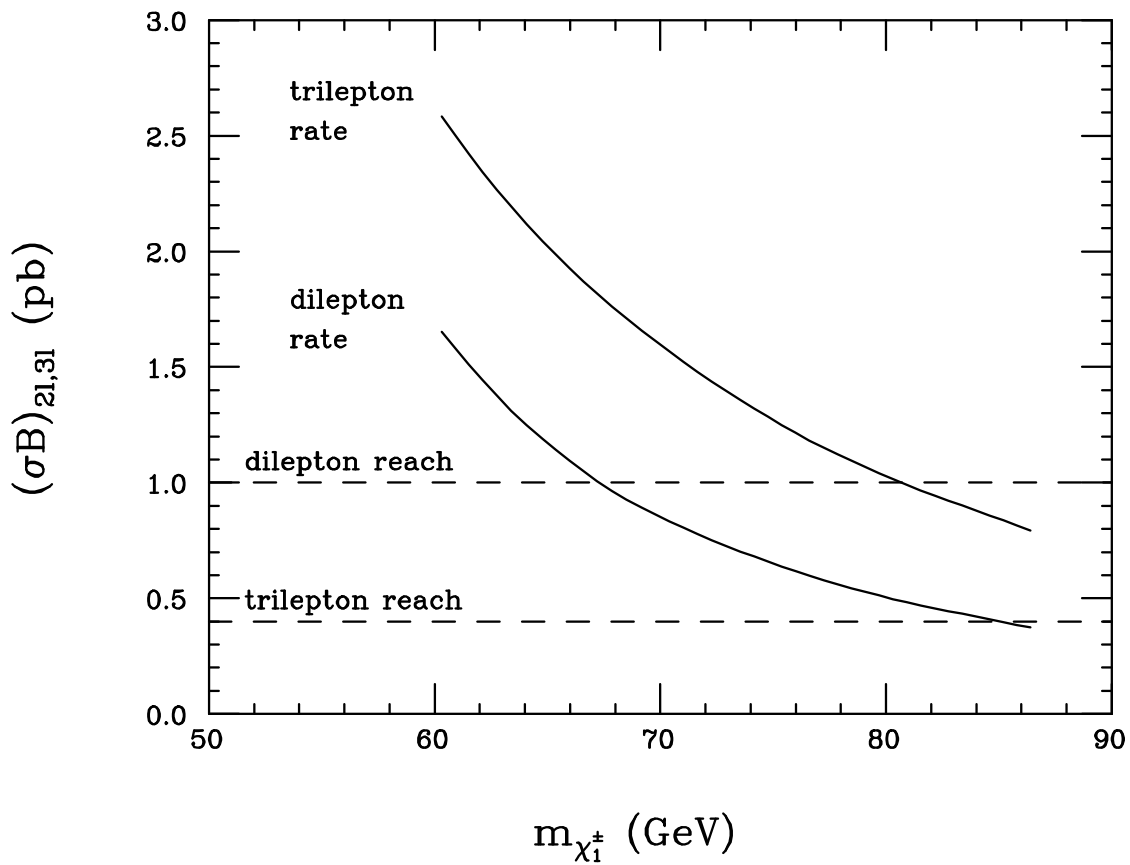}
\caption{The dilepton and trilepton rates at the Tevatron versus the chargino
mass originating from neutralino and chargino production. The indicated reaches
are expected with $100\,{\rm pb}^{-1}$ of accumulated data.}
\label{di-tri}
\end{figure}

At the present-day Tevatron the strongly interacting gluino and squarks are not
accessible ($m_{\tilde g,\tilde q}\gsim300\GeV$), although they should be
easily detectable at the LHC (note that top-squarks ($\tilde t_{1,2}$) are
considerably split relative to the other squarks). On the other hand, the
weakly interacting neutralinos and charginos are quite reachable in this model
via the trilepton signal in $p\bar p\to\chi^0_2\chi^\pm_1X$ \cite{trileptons}
and the dilepton signal in $p\bar p\to\chi^+_1\chi^-_1X$ \cite{di-tri}. The
chargino ($\chi^\pm_1$) branching ratio into leptons ($e+\mu$) is $\approx0.5$,
whereas that into jets is $\approx0.25$, for all allowed points in parameter
space. Also, the neutralino ($\chi^0_2$) decays exclusively to dileptons
because of the dominant two-body decay mode
$\chi^0_2\to\tilde\ell_R^\pm\ell^\mp$. Therefore, the trilepton and dilepton
signals are nearly maximized. The corresponding rates are shown in
Fig.~\ref{di-tri}. The expected experimental sensitivities with $100\ipb$ of
accumulated data (end of 1995) are indicated by the dashed lines,\footnote{The
sensitivities are actually chargino-mass
dependent, following a curve shaped similarly to the signal and which
asymptotes to the indicated dashed lines. With the CDF data from Run IA
($\approx20\ipb$), this asymptote is at $\approx2\pb$ \cite{Kato}.} which shows
guaranteed discovery of the chargino via the trilepton mode.
Right-handed slepton pair-production at the Tevatron ($p\bar p\to\tilde
\ell^+_R\tilde \ell^-_RX$) will produce exclusively dileptons since
$B(\tilde\ell_R\to\ell\chi^0_1)=1$. However, the cross section is not
large \cite{BaerSleptons}: $1\,(0.1)\pb$ for $m_{\tilde\ell_R}=45\,(50)\GeV$
and the backgrounds are not small. With $100\ipb$ and $10\%$ detection
efficiency the lower end of the allowed range could be explored.

At LEPII, starting in 1996 with a center-of-mass energy of $\sqrt{s}=180\GeV$,
it should be possible to observe the lightest Higgs boson, the lightest
chargino, and the right-handed sleptons. The Higgs-boson coupling to gauge
bosons is indistinguishable from the Standard Model prediction
($\sin^2(\alpha-\beta)=0.996-0.998$), although its branching ratio into $b\bar
b$ may be eroded somewhat for the lightest allowed masses, because of the
supersymmetric decay channel with $B(h\to\chi^0_1\chi^0_1)<0.18$. The reach of
LEPII for Higgs masses is estimated at $\sqrt{s}-95=85\GeV$ \cite{Sopczak}, and
thus an increase of center-of-mass energy to $\sqrt{s}=190\GeV$ would allow
full discovery potential for the Higgs boson. The charginos would be pair
produced ($e^+e^-\to\chi^+_1\chi^-_1$) and in the preferred ``mixed" decay mode
(1 lepton + 2 jets) one chargino decays leptonically and the other one
hadronically. Since neither of the chargino branching fractions is suppressed,
we find a cross section into the mixed mode as large as $2.1\pb$ and decreasing
down to $0.62\pb$ at the upper end of the allowed interval. This signal should
be readily detectable \cite{Grivaz}. One would also produce $\chi^0_1\chi^0_2$
with a cross section from $1.4\pb$ down to $0.6\pb$, which could only be
detected via the dilepton mode (since $B(\chi^0_2\to 2j)\approx0$). The
right-handed selectrons (smuons) have a pair-production cross section exceeding
 $2\pb$ ($0.9\pb$) and should be easily detectable over the $WW$
background \cite{Grivaz}.

One can also test the model via rare processes. The prediction for $B(b\to
s\gamma)$ varies widely with the supersymmetric spectrum \cite{bsgamma} and
is subject to large QCD uncertainties, accounting for these as
described in Ref.~\cite{Buras} we get a range: $(4.2\to5.3)\times10^{-4}$
for the lower end and $(3.9\to5.1)\times10^{-4}$ for the upper end of the
spectrum. These predictions are in fair agreement with the present
experimentally allowed range of $(1-4)\times10^{-4}$ \cite{CLEO}. For the
supersymmetric contribution to the anomalous magnetic moment of the muon
\cite{g-2} we get $a^{\rm susy}_\mu=(-2.4\to-1.7)\times10^{-9}$, which is not
in conflict with present experimental limits but could be easily observable at
the new E821 Brookhaven experiment which aims at a sensitivity of
$0.4\times10^{-9}$. We have also computed the supersymmetric contribution to
the ratio $R_b=\Gamma(Z\to b\bar b)/\Gamma(Z\to{\rm hadrons})$ \cite{Rb} which
is measured at LEP. We find $R^{\rm susy}_b=(4.4\to3.2)\times10^{-4}$, which
would require a four-fold increase in the experimental sensitivity to be
observable.

Finally we discuss the cosmological implications of our model. The relic
abundance of the lightest neutralino has been calculated following the methods
of Ref.~\cite{LNYdm} and determined to be $\Omega_\chi h^2\approx0.025$ with a
dip towards the end of the allowed range when the $Z$ pole is encountered in
the neutralino annihilation. Such small cold dark matter density would be of
interest in models of the Universe with a significant cosmological
constant \cite{CC}. Such neutralinos would populate the galactic halo and could
be detected in cryogenic detectors. The calculated rate \cite{lspd} in the
soon-to-be-operational Stanford Germanium detector is enhanced for
$m_{\chi^0_1}\approx{1\over2}m_{\rm Ge}\approx37\GeV$, reaching a maximum of
$R=0.06\,{\rm (events/day/kg)}$. With an expected sensitivity of $R=0.1$ (or
perhaps $R=0.01$ eventually) this discovery channel does not appear especially
promising.

In conclusion, we have presented a string-derived no-scale supergravity model
where all parameters are calculated from first principles and where all sectors
of the model are correlated for the first time. This model can be described
in terms of a single parameter which is phenomenologically strongly restricted,
so much that the lighter charginos, neutralinos, sleptons, and Higgs boson
would become observable at the Tevatron and LEPII in the very near future.

\section*{Acknowledgments}
We would like to thank T. Kamon and J.~T.~White for useful discussions.
This work has been supported in part by DOE grant DE-FG05-91-ER-40633.

\clearpage

\clearpage

\clearpage

\end{document}